# Global research collaboration: Networks and partners in South-East Asia

**Richard Woolley[1], Nicolas Robinson-Garcia[2] and Rodrigo Costas[3,4]**

[1]INGENIO (CSIC-UPV), Universitat Politècnica de Valéncia, Valencia (Spain)
[2]School of Public Policy, Georgia Institute of Technology, Atlanta (United States)
[3]CWTS, Leiden University, Leiden (The Netherlands)
[4]DST-NRF Centre of Excellence in Scientometrics and Science, Technology and Innovation Policy, Stellenbosch University, South Africa

## Abstract

This is an empirical paper that addresses the role of bilateral and multilateral international co-authorships in the six leading science systems among the ASEAN group of countries (ASEAN6). The paper highlights the different ways that bilateral and multilateral co-authorships structure global networks and the collaborations of the ASEAN6. The paper looks at the influence of the collaboration styles of major collaborating countries of the ASEAN6, particularly the USA and Japan. It also highlights the role of bilateral and multilateral co-authorships in the production of knowledge in the leading specialisations of the ASEAN6. The discussion section offers some tentative explanations for major dynamics evident in the results and summarises the next steps in this research.

## 1. Introduction

In the 'flat earth' ideal of global science networks, access and contribution to science is no longer structured by zones of inclusion and exclusion. Geographic locations and local resources are viewed as secondary to international connectedness. Raising scientific quality is viewed as contingent on plugging into global networks, which are assumed to underpin an R&I driven mode of socio-economic development. An alternative interpretation of the globalized organization of science sees global networks as a perpetuation of asymmetric relations of power and control over the scientific agenda. In this view global networks export the research agenda of rich, successful countries, enrolling research groups in other locations. Research quality becomes disconnected from local societal relevance.

Debate over these polar visions of the role of scientific networks is no closer to being settled. Even more basic, the involvement of developing science and research systems in global networks, and the evolution of these networks over time, remains unclear with regards to many countries and regions. The perpetuation of this research gap provides the main rationale for this paper. The empirical contribution of the paper is to provide clarity about the international collaboration patterns of a group of six emerging science research systems in south-east Asian. The paper also makes a conceptual contribution to studies of international collaboration by distinguishing between bi-lateral and multi-lateral collaboration. The paper uses this distinction to produce new insights into how international collaborations are structured.

The paper primarily reports new empirical knowledge. The empirical work is conducted using bibliometric data and analyses, limiting the interpretations that can be made of the results. Nevertheless, possible explanations for the different structures of international collaboration described are discussed, including explanation driven by knowledge specialisations, patterns of research training, and institutional factors including funding rules.

## 2. Global collaboration in science

This section is divided into two parts. The first part reviews selected literature on global collaboration in science. The second part introduces an analytic distinction between a bi-lateral mode of international collaboration and multi-lateral, or more properly networked, mode of international collaboration.

### 2.1 Relevant previous research

A continuous growth in the number of international co-authorships of scientific papers has been observed over the last decade (Allen 2017). The share of scientific papers that are international co-authorships has more than double in twenty years (Wagner et al. 2015). This rise in international co-authorships has been equivalent to all the growth in output of advance science systems in recent years (Adams 2013). According to Allen (2017), in the life science in particular there has also been an increase in the number of authors per paper.

In terms of the global co-authorship network, an initial core network of Western European nations and the USA (Leydesdorff et al. 2013), expanded to an extended leading group of around 64 countries by 2005 and 114 countries by 2011 (Wagner et al. 2015). There are now many more countries participating in global co-authorships than was the case twenty years earlier (Bornmann et al. 2015). The density of the network has also increased, a 'many more connections have been forged by more partners' (Wagner et al. 2015: 6). From the perspective of developing research systems, a significant proportion of the expansion in international co-authorships can be explained by the broadening of country participation in the production of scientific outputs. This participation is facilitated by global interconnectedness in communication and travel, and driven by a range of personal and institutional advantages of collaborating with excellent researchers and high quality research organisations in leading countries, and by the increasing investment and capacity within developing systems themselves (Wagner and Leydesdorff 2005; Wagner et al. 2015).

Wagner and colleagues (2017: 1645) also found that the number of co-author countries involved in a publication is significantly correlated with higher citation impact in some scientific disciplines, including virology, soil science, seismology, and astrophysics. To the extent that citation impact can be considered a relevant measure, this suggests that there may be improvement in scientific quality associated with the extension of international co-authorship networks.

Scientific publications from the ASEAN nations has been increasing at a faster rate than for the global system overall, slowly but steadily increasing their small share of global scientific output (Nguyen and Pham 2011). The six largest science producing countries are Indonesia, Malaysia, The Philippines, Singapore, Thailand and Vietnam. These countries collaborate most strongly on international publications with the Japan, USA and the EU block as a whole (Hassan et al. 2012). Internal collaboration within the Asia-Pacific region, (ASEAN plus Australia, China, Japan, South Korea, New Zealand, and Taiwan) has also been shown to be intensifying in the past decade (Haustein et al 2011). However, whilst significant research has highlighted which countries are the most important international collaboration partners of the major ASEAN knowledge producers, there has been no previous investigation of whether international collaboration patterns in the region tend to involve just one international partner or to include multiple different international co-authors.

## 2.2. The network mode of international collaboration in science

In the empirical analyses of the six largest ASEAN science producers (ASEAN6) that follow, we introduce a distinction between bi-lateral international research collaboration and multi-lateral international research collaboration. In network theory terms we are introducing a distinction between dyadic collaborations and those that are, at least, triadic collaborations. In network analysis, dyadic collaborations can be measured in terms of their balance in terms of power, reciprocity, mutuality and other elements. A triadic network has a minimum of four potential sets of dyadic relations, as shown in Figure 1. Once there are a minimum of three nodes, then measures of centrality, brokering, and bridging can also be used.

Figure 1 Dyads and triads

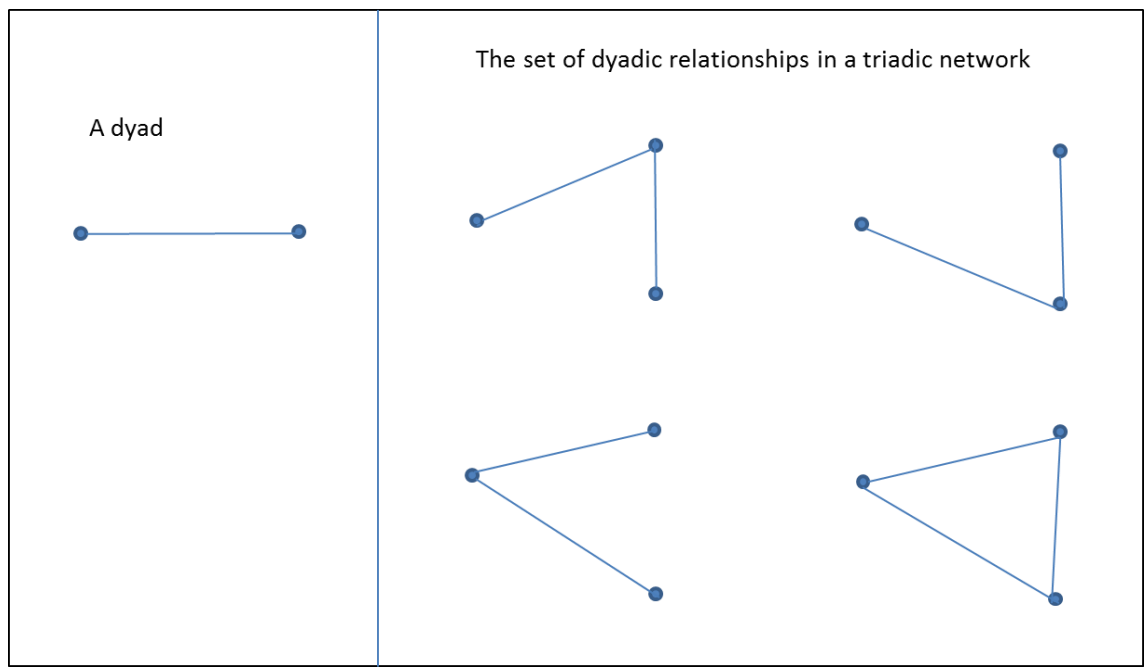

As was described in the previous section international collaborations in science are typically measured through co-authorships of scientific journal articles. An international collaboration is defined by the author address field, with a minimum of two different countries constituting an international collaboration. In the case of two authors in two different countries, these authors must know each other. If there are multiple authors on one or both sides of the international divide, then at least one of the pairs of authors spanning this divide must know each other.

However, in the case of three (or more) authors in three different countries this is not necessarily the case. Provided at least one author in country A knows at least one author in country B and in country C, then there is no necessary reason why a pair of authors in countries B and C must know each other.[1] In a triadic network there are three possible such combinations of mediating author linking two dyads, plus a full set of three dyadic relations.

In this paper we refer to two country (dyadic) international research co-authorships as bilateral international research co-authorships (BIRCs) and three country (triadic), or more complex, international research co-authorships as multi-lateral international research co-authorships (MIRCs). Our rationale for doing this is that our understanding of the complex picture of global science collaboration will be strengthened by knowing to what degree overall patterns are structured by bilateral relationships and by more complex networks. Our hypothesis is that the effects of BIRCs and MIRCs on the structuring of global research collaboration will be different. For convenience sake, we also refer to BIRCs as the *partnership mode* of international research collaboration and MIRCs as the *network mode* of international research collaboration.

## 3. Data and methods

In this paper we analyzed a total of 28,429,592 publications indexed in the Web of Science and published between 1980 and 2015. We used the CWTS in-house database and selected only articles and reviews, removing non-research document types such as editorial material, notes, news items, meeting abstracts, etc. For each publication, we calculated the number of collaborating countries. Following the model developed by de Lange and Glänzel (1997) and Glänzel and de Lange (1997), we define three types of international collaboration:

- No collaboration. All researchers authoring a paper are affiliated to institutions from the same country.
- Bilateral collaboration (BIRCs). Researchers authoring a paper are affiliated to institutions from two different countries.
- Multilateral collaboration (MIRCs). Researchers authoring a paper are affiliated to institutions from more than two different countries.

We then focused our analysis on six ASEAN countries: Thailand, Singapore, Malaysia, Philippines, Vietnam and Indonesia (ASEAN6). Table 1 shows the total number of papers produced by each of these countries and their shares by type of collaboration. As observed, while Malaysia and Thailand present similar shares of bilateral and multilateral collaboration, Indonesia, Philippines and Vietnam show a complete different pattern with bilateral collaboration being more prominent. Finally, Singapore shows a different pattern with the share of multilateral collaboration almost doubling the share with bilateral collaboration.

For both bilateral and multilateral collaboration, for the global system and the ASEAN6 sub-system, we developed coauthorshp networks using VOSviewer (van Eck & Waltman 2010). We used the LinLog layout technique and modularity clustering (Newman 2004; Noack 2007, 2009) to visualize the networks. The rest of the analysis is of a descriptive nature.

---

[1] In fact, one author in Country A may know one author in country B and another author in country A may know one author in Country C, provided the two authors in country A know each other.

Table A. Overview of the scientific production of ASEAN6, by co-authorship type, 1980-2015

| ASEAN6 | # Pubs | % Bilateral collaboration | % Multilateral collaboration |
|---|---|---|---|
| Indonesia | 19,038 | 54,2% | 26,9% |
| Malaysia | 82,452 | 12,5% | 13,1% |
| Philippines | 16,769 | 61,6% | 22,6% |
| Singapore | 149,657 | 6,9% | 13,2% |
| Thailand | 77,383 | 13,3% | 15,5% |
| Vietnam | 20,862 | 49,5% | 24,0% |

# 4. Results

This section presents our results in four parts. First, we show the global trend in bi-lateral international research collaborations (BIRCs) and multi-lateral international research collaborations (MIRCs). We then map the BIRC and MIRC networks at the global level. Second, we show these same results for the ASEAN6 nations. Third, we move to the individual country level of our six focus countries, to look at the major collaborating countries for both BIRCs and MIRCs. Finally, we compare the rankings of subject areas of the BIRC and MIRC co-authored papers.

## 4.1 International co-authorships

The majority of scientific papers are produced by single authors, or multiple authors based in the same country. However, the share of international co-authorships has been growing over time (Figure 2).

Figure 2 Share of international collaborations, by type, 1980-2015

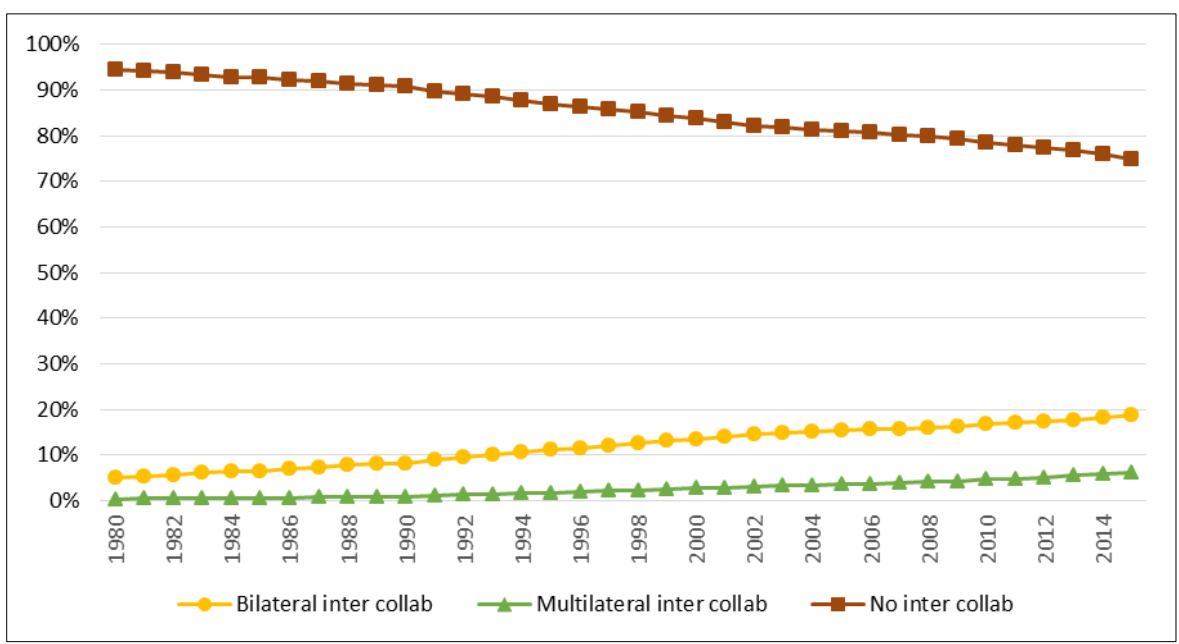

The share of scientific research papers that are produced by international collaborations grew from around 10% in 1990 to around 20% in 2005. This share remained relative constant, for around five years before increasing to around 25% of all papers in 2015. It appears that growth in the share of BIRCs drove growth in international collaborations until this most recent phase, when growth in international collaborations was driven by a faster rate of growth in MIRCs.

These data show that the dynamics of international co-authorships are undergoing a transformation in terms of the numbers of different international partners that are combining to produce scientific outputs. Whilst overall international co-authorships continue to grow as a share of all scientific articles, the proportion of these papers which are the outputs of MIRCs is growing more strongly in recent years. Our initial hypothesis was that BIRCs and MIRCs would play different roles in the overall structuring of global science networks. Figure 3 maps the global network of BIRCs and MIRCs.

Figure 3 Global BIRCs (above) and MIRCs (below) networks, 1980-2015, minimum 1000 papers

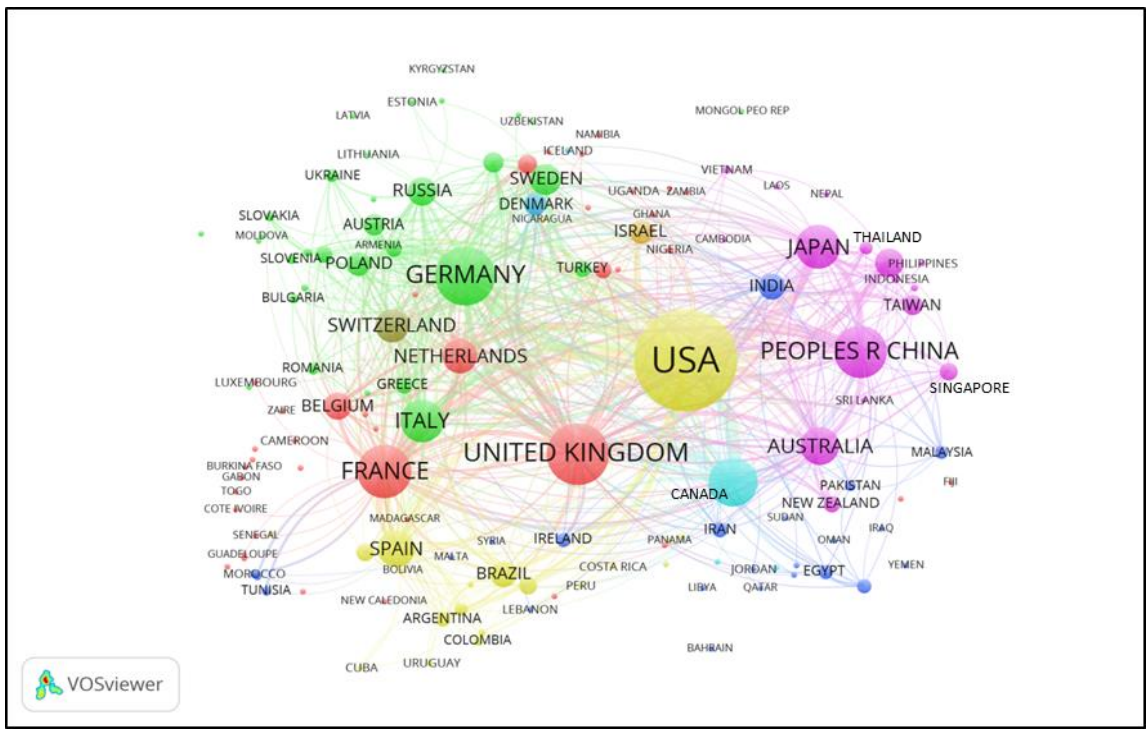

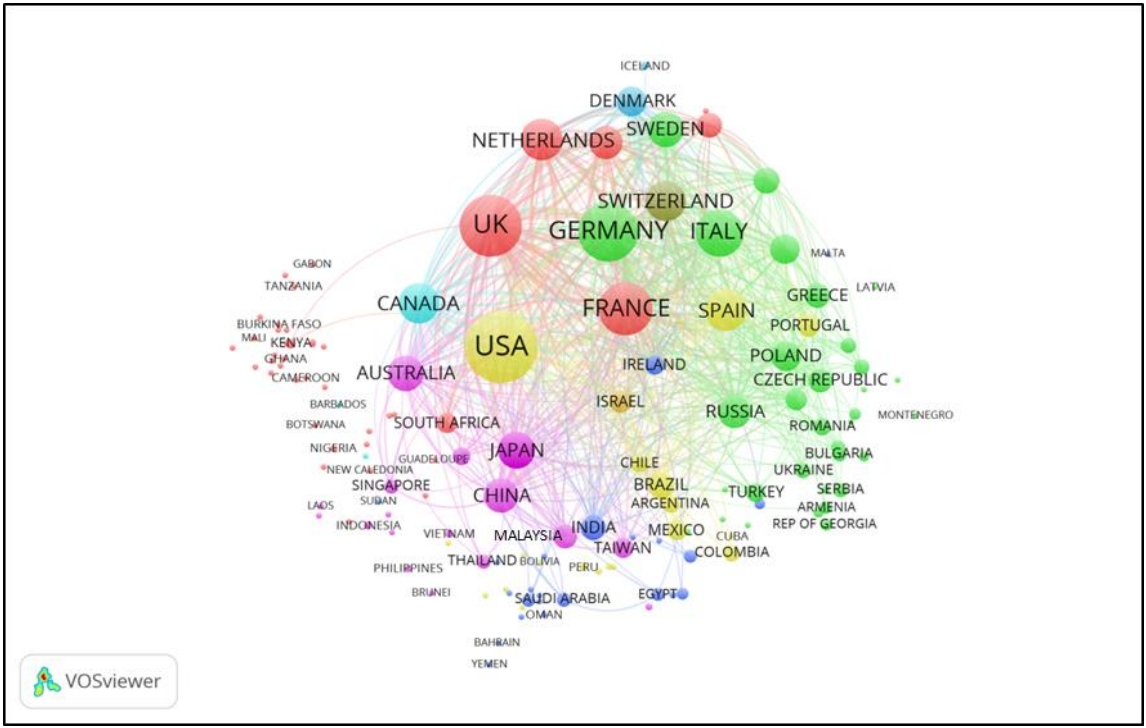

The USA and the UK are in prominent central positions in the global BIRCs network, between western European, Scandinavian and Asia Pacific groupings. In the MIRCs network the UK is more closely linked to northern and western European neighbours. The USA is less central and dominant, seemingly more closely embedded with Asia Pacific regional network. In both networks Germany also bridges into southern and eastern Europe, while an Anglophone grouping of the USA, UK, Canada and, to a lesser extent Australia, link together quite centrally.

The ASEAN6 countries are somewhat differently position in the BIRCs and MIRCs networks. In the BIRCs network, Indonesia, the Philippines, Singapore and Thailand are closely tied to Japan and China. In the MIRCs network, the Indonesia, Philippines and Vietnam in particular seem relative disconnected from the global co-authorships. Singapore remains relatively well connected to at least four strong partners, whilst Malaysia shifts somewhat and appears more embedded in the MIRCs network.

In terms of our starting hypothesis that BIRCs and MIRCs would contribute to the structuring of the global co-authorships network in different ways, the initial evidence appears to provide sufficient justification for the development of more sophisticated analyses in the future.

## 4.2 ASEAN6 international co-authorships

A large share of scientific papers produced in ASEAN6 countries are the outputs of single authors, or multiple authors based in the same country. However, the share of international co-authorships has been growing over time and now constitutes a majority of ASEAN6 scientific output (Figure 4).

Figure 4 Share of international collaborations, ASEAN6 by type, 1980-2015

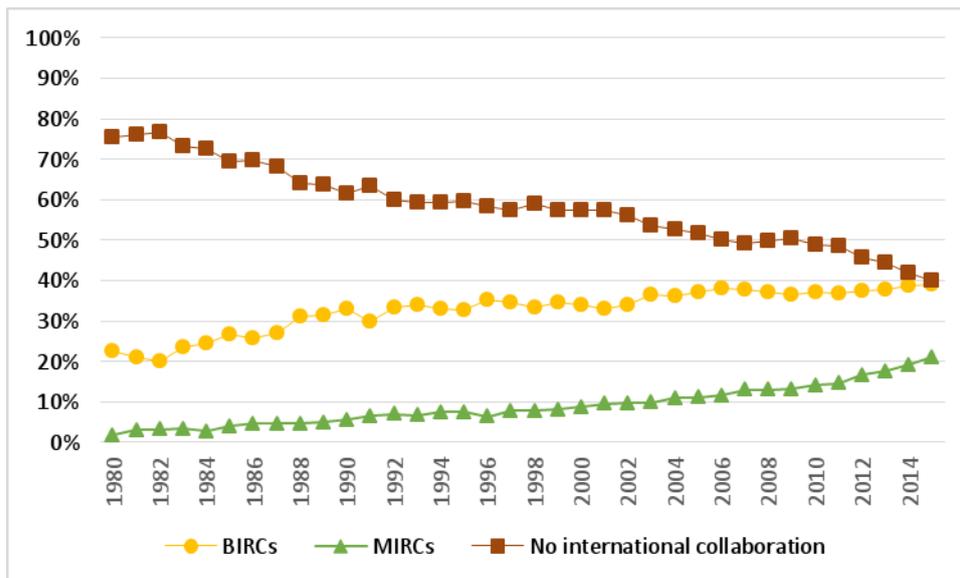

ASEAN6 countries are more reliant on international co-authorships compared to level found in the global system as a whole. In the most recent decade MIRCs have been driving the growth in international collaborations, reaching 20% of all output or the equivalent of 33% of all international collaborations in 2015. There is, of course, considerable variation between ASEAN6 countries in terms of rate of international co-authorships, ranging from around 50 % (Malaysia, Thailand) to more than 80% (Indonesia) of all publications in 2015. In Singapore, the international co-publication ratio has grown from 20% to around 62% of all scientific papers since 1990. Whether there is also variation at the individual country level in terms of the balance of BIRCs and MIRCs within overall international collaboration is assessed in Section 4.3. However, next we compare the BIRCs and MIRCs networks for the ASEAN6 (Figure 5).

The MIRCs network is significantly more extended than the BIRCs network, at the threshold level for inclusion of 150 scientific papers. The BIRCs network includes 48 countries while the MIRCs network includes 98 countries. Singapore, Thailand and Malaysia are the most prominent countries in both the BIRCs and MIRCs ASEAN6 co-authorship networks. The BIRCs network has something of a three-pole structure, with Malaysia and Thailand more central than Singapore in both networks. In contrast, while Singapore is the largest participant in both the ASEAN6 international co-authorships networks it is less central. In both networks the Singapore pole is positioned close to China and the USA. At the opposite side of the BIRCs network is Japan, in between the other five countries in the ASEAN6.

Japan remains central to the MIRCs network in relation to Indonesia, the Philippines, Thailand and Vietnam. Malaysia on the other hand remains more evenly balanced between Japan and the UK collaborations, with India also a more prominent partner. At the same time, Malaysia is the main linkage point to both networks for a small number of members of the Organisation of Islamic Cooperation (OIC). Australia, Taiwan and the UK are all relatively central in both networks, bridging between the three main ASEAN6 poles. South Korea is also more broadly involved in the MIRCs network than is the case for BIRCs network. It is also noticeable that in the MIRCs network Singapore has important co-authorship relationships with Canada, Germany and Sweden.

Figure 5 ASEAN6 BIRCs (above) and MIRCs (below) networks, 1980-2015, minimum 150 papers

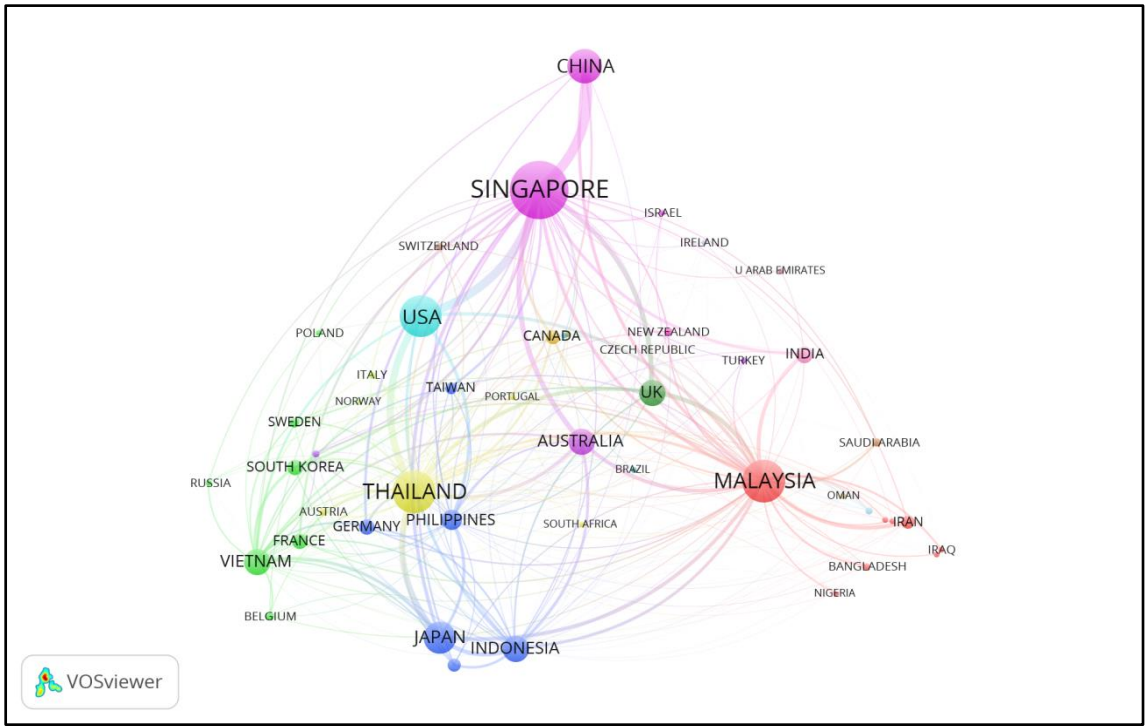

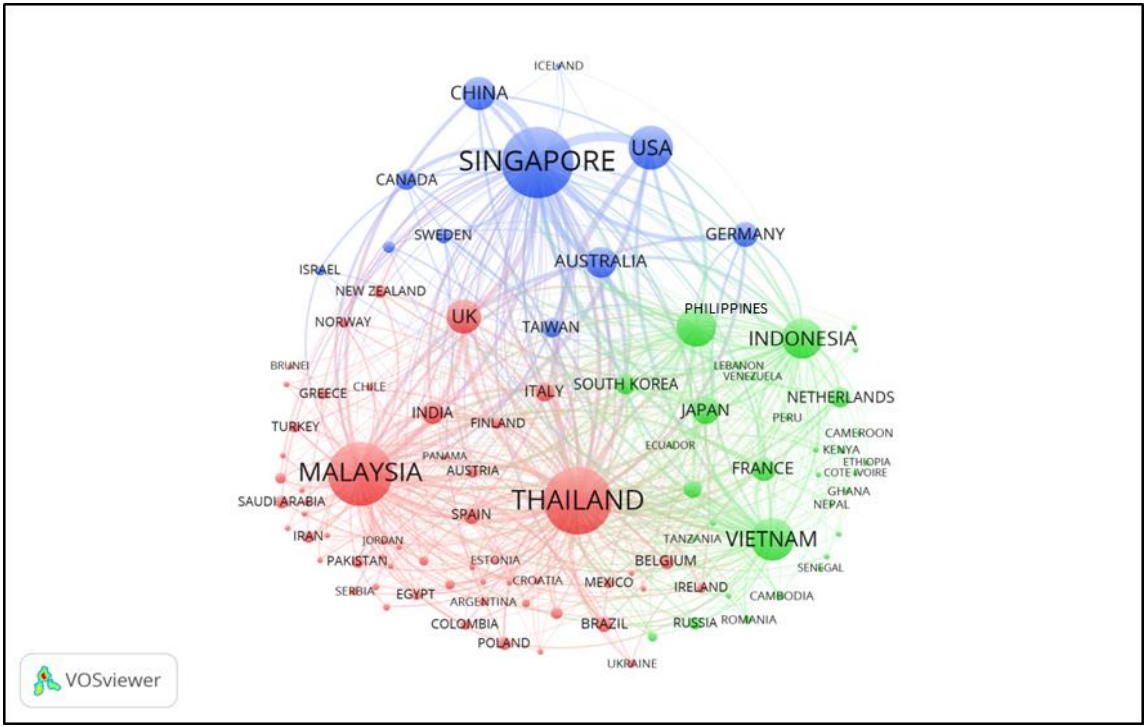

In terms of our starting hypothesis that BIRCs and MIRCs would contribute to the overall structuring of co-authorship networks in different ways, the initial evidence for the ASEAN6 group of countries appears to also support this contention. Having also established the apparent emergence of MIRCs as the driver of growth in co-authorships at both the global and the ASEAN6 level, in the following section we look at data for individual countries, seeking to encounter variation within our data that may open up discussions and contribute to building explanations regarding the emergence and growth of BIRCs and MIRCs in different contexts.

## 4.3 International co-authorship networks of the ASEAN6

This section looks at the international co-authorships of each of the ASEAN6 countries in turn. All ASEAN6 countries have a strong degree of international co-authorships as a proportion of their total output. However, there is also considerable variation among the group (Figure 6).

Figure 6 International co-publication rates, selected SE Asian countries (%)

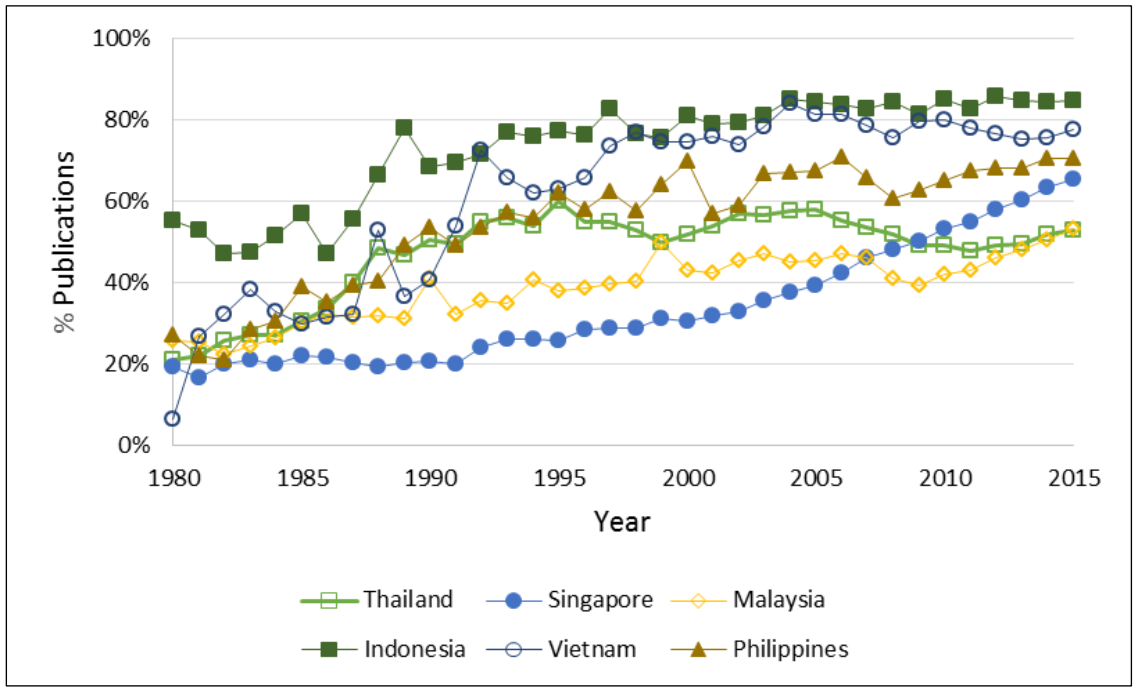

The international co-authorship trend of has already reached 50% by the early 1990s for four of the ASEAN6. In the cases of Indonesia and Vietnam the international co-authorship rate has been 80% of total output for around 20 years. The ratio has also been very high (60-70%) across this period for the Philippines. Thailand has maintained an international co-authorship rate between 50% and 60% over this time. Malaysia reached an international co-authorship rate of 50% around ten years later in 1999 and has maintained in the range 40-50% since, rising above 50% again in the most recent years of the series. Singapore has a different trajectory, with international co-authorships growing steadily from around 30% in 2000 to over 60% by 2015. Table 1 summarises the major locations of co-authors for each of the ASEAN6 over the period.

Table 1 Main international collaborating partners, ASEAN 6, all articles & reviews 1980-2015

| Malaysia | | | Indonesia | | | Vietnam | | |
|---|---|---|---|---|---|---|---|---|
| Country | Pubs | % all pubs | Country | Pubs | % all pubs | Country | Pubs | % all pubs |
| UK | 6236 | 7,56% | JAPAN | 3936 | 20,67% | USA | 2660 | 12,75% |
| USA | 4764 | 5,78% | USA | 3224 | 16,93% | JAPAN | 2506 | 12,01% |
| AUSTRALIA | 4456 | 5,40% | AUSTRALIA | 2690 | 14,13% | FRANCE | 2237 | 10,72% |
| INDIA | 4181 | 5,07% | NETHERLANDS | 1890 | 9,93% | SOUTH KOREA | 2010 | 9,63% |
| JAPAN | 3942 | 4,78% | UK | 1604 | 8,43% | UK | 1719 | 8,24% |
| CHINA | 3532 | 4,28% | *MALAYSIA* | *1458* | *7,66%* | GERMANY | 1593 | 7,64% |
| IRAN | 3459 | 4,20% | GERMANY | 1251 | 6,57% | AUSTRALIA | 1473 | 7,06% |
| SAUDI ARABIA | 2057 | 2,49% | FRANCE | 1006 | 5,28% | CHINA | 1239 | 5,94% |
| *SINGAPORE* | *1971* | *2,39%* | *THAILAND* | *784* | *4,12%* | NETHERLANDS | 1098 | 5,26% |
| *THAILAND* | *1735* | *2,10%* | CHINA | 700 | 3,68% | *THAILAND* | *861* | *4,13%* |
| **The Philippines** | | | **Singapore** | | | **Thailand** | | |
| Country | Pubs | % all pubs | Country | Pubs | % all pubs | Country | Pubs | % all pubs |
| USA | 3616 | 21,56% | USA | 22484 | 15,02% | USA | 13803 | 17,84% |
| JAPAN | 2304 | 13,74% | CHINA | 22047 | 14,73% | JAPAN | 8257 | 10,67% |
| AUSTRALIA | 1363 | 8,13% | UK | 8719 | 5,83% | UK | 5895 | 7,62% |
| CHINA | 1125 | 6,71% | AUSTRALIA | 8425 | 5,63% | AUSTRALIA | 4206 | 5,44% |
| UK | 989 | 5,90% | JAPAN | 4071 | 2,72% | CHINA | 3132 | 4,05% |
| INDIA | 831 | 4,96% | CANADA | 3794 | 2,54% | FRANCE | 2622 | 3,39% |
| GERMANY | 767 | 4,57% | GERMANY | 3726 | 2,49% | GERMANY | 2521 | 3,26% |
| *THAILAND* | *751* | *4,48%* | INDIA | 2982 | 1,99% | CANADA | 1790 | 2,31% |
| SOUTH KOREA | 681 | 4,06% | FRANCE | 2731 | 1,82% | *MALAYSIA* | *1735* | *2,24%* |
| TAIWAN | 669 | 3,99% | SOUTH KOREA | 2588 | 1,73% | SOUTH KOREA | 1615 | 2,09% |

Authors based in the USA are the most frequent international collaborators for ASEAN6 authors in four countries, and the second most frequent collaborators in the other two. Japan is also a very frequent collaborator particularly with Indonesia and Vietnam, but also with the Philippines and Thailand. The two main exceptions to the prominence of the US and Japan are the UK with Malaysia and China with Singapore. Australia is a consistently strong collaborator across the ASEAN6. Germany and France are the most prominent co-author countries from continental Europe. Malaysia's connection to countries in the Middle East region is noticeable, as are the prominence of France and the Netherlands as partners for Vietnam and Indonesia respectively.

### 4.3.1 International co-authorships: Indonesia
Indonesian authors have been historically strongly involved in international co-authorship and continue to be so. BIRCs have made up more than half of Indonesia's total output for most of the period 1908-2015, with the rise of MIRCs progressively replacing sole authored papers as a share of total output (Figure 7).

Figure 7 International co-authorships by type, Indonesia 1980-2015

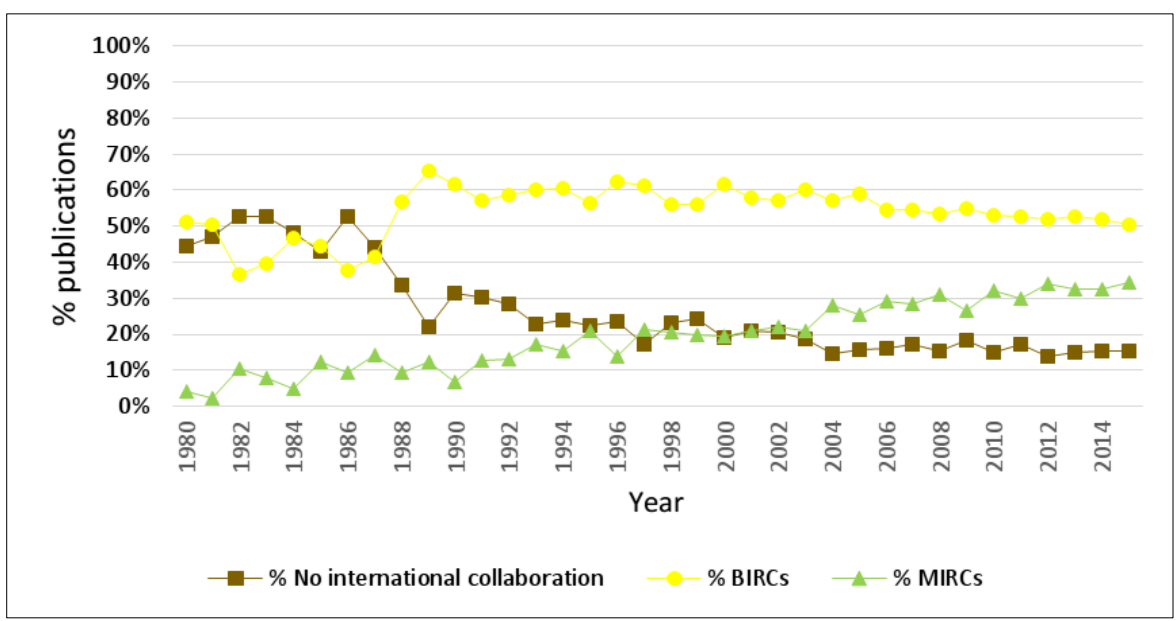

In 2015, BIRCs made up around 50% of Indonesia's scientific output, while MIRCs accounted for around 33%. In terms of co-authorship partners, Japan, the USA and Australia were the three most common locations of co-authors in the Indonesian case.

Table 2 International co-authorship countries, by type, Indonesia 1980-2015

| INDONESIA | | | |
|---|---|---|---|
| BIRCs | | MIRCs | |
| country | pubs | country | pubs |
| JAPAN | 2764 | USA | 1816 |
| USA | 1360 | AUSTRALIA | 1296 |
| AUSTRALIA | 1344 | UK | 1099 |
| NETHERLANDS | 1096 | JAPAN | 1095 |
| MALAYSIA | 717 | NETHERLANDS | 778 |
| GERMANY | 584 | MALAYSIA | 739 |
| UK | 486 | THAILAND | 677 |
| FRANCE | 418 | GERMANY | 645 |
| SOUTH KOREA | 229 | PR CHINA | 640 |
| TAIWAN | 195 | FRANCE | 582 |

Japan is the main location of co-authors for partnership mode collaborations, while the USA is the main location of co-authors for network mode collaborations. Authors based in the USA, UK and Germany participated in more network mode than partnership mode collaborations, while Australia and Malaysia are relatively balanced between the two. Authors based in Japan and the Netherlands have more partnership mode than network mode collaborations with Indonesia-based co-authors.

### 4.3.2 International co-authorships: Malaysia

Malaysia-based authors have published a majority of their scientific articles in collaborations with international co-authors in only the final two years of the analysis period. BIRCs have made up around one-third of Malaysia's total output since the mid-1990s. MIRCs have been increasing of a share of scientific output across the period, with a more rapid growth evident in the past 5-7 years (Figure 8).

Figure 8 International co-authorships by type, Malaysia 1980-2015

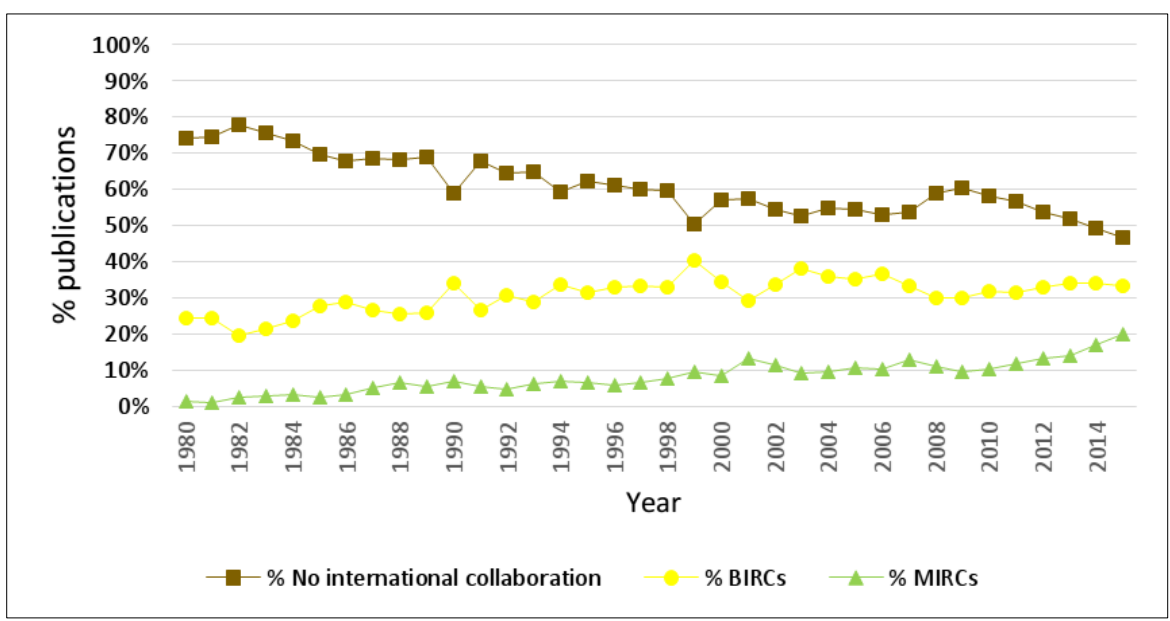

In 2015, BIRCs made up around 33% of Malaysia's scientific output, while MIRCs accounted for around 20%. In terms of co-authorship partners, the UK, the USA and Australia were the three most common locations of co-authors in the Malaysian case (Table 3).

Table 3 International co-authorship countries, by type, Malaysia 1980-2015

| MALAYSIA | | | |
|---|---|---|---|
| BIRCs | | MIRCs | |
| country | pubs | country | pubs |
| UK | 3196 | UK | 2906 |
| IRAN | 2461 | USA | 2863 |
| INDIA | 2353 | AUSTRALIA | 2101 |
| AUSTRALIA | 2266 | PR CHINA | 1839 |
| JAPAN | 2181 | INDIA | 1764 |
| USA | 1858 | JAPAN | 1716 |
| PR CHINA | 1681 | SAUDI ARABIA | 1250 |
| SINGAPORE | 822 | THAILAND | 1222 |
| SAUDI ARABIA | 797 | GERMANY | 1166 |
| PAKISTAN | 768 | SINGAPORE | 1126 |

The UK is the main location of co-authors for both partnership mode and network collaborations of Malaysia-based authors. Iran is the second most common location of co-authors for partnership mode collaborations, although Iran-based authors are less prominent, although still significant participants, in network mode collaborations (n=982). India, Australia and Japan are other prominent collaborators, more strongly in partnership mode than network mode. On the other hand, the USA, China, Saudi Arabia, Thailand, Germany and Singapore are all prominent collaborators which are more strongly involved in network mode than partnership mode collaboration.

### 4.3.3 International co-authorships: the Philippines

Philippines-based authors have been publishing a majority of their scientific articles in collaborations with international co-authors since about 1990. BIRCs averaged around 45% of scientific output from the Philippines from 1990 to around 2005. Over the past decade, strong growth in MIRCs has led to a situation where the BIRCs, MIRCs, and no international collaboration, make up approximately equal shares of the Philippines' total output (Figure 9).

Figure 9 International co-authorships by type, the Philippines 1980-2015

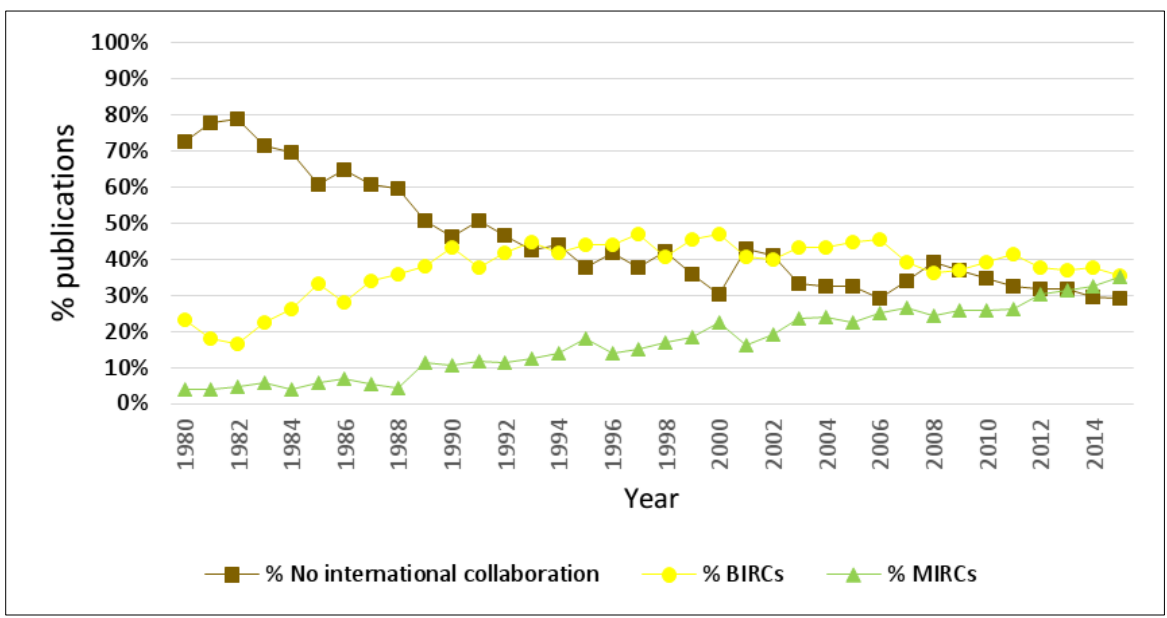

In 2015, BIRCs and MIRCs combined made up around 70% of the Philippines' scientific output. In terms of co-authorship partners, the USA, Japan and Australia were the three most common locations of co-authors in the Philippines case (Table 4).

Table 4 International co-authorship countries, by type, the Philippines 1980-2015

| PHILIPPINES | | | |
|---|---|---|---|
| **BIRCs** | | **MIRCs** | |
| **country** | **pubs** | **country** | **pubs** |
| USA | 1761 | USA | 1759 |
| JAPAN | 1421 | JAPAN | 831 |
| AUSTRALIA | 529 | PR CHINA | 825 |
| PEOPLES R CHINA | 293 | AUSTRALIA | 800 |
| GERMANY | 271 | UK | 719 |
| UK | 261 | THAILAND | 671 |
| TAIWAN | 254 | INDIA | 670 |
| SOUTH KOREA | 230 | GERMANY | 487 |
| NETHERLANDS | 163 | MALAYSIA | 458 |
| INDIA | 149 | SOUTH KOREA | 449 |

The USA is the main location of co-authors for both partnership mode and network collaborations of Philippines-based authors, with Japan second for both modes. USA co-authorships are evenly split between partnership and network modes, whilst Japan is strongly biased toward the partnership mode. China is a prominent collaborator country that is much more involved in network mode collaborations. The remainder of the prominent co-author countries are, likewise, all involved more strongly in the network mode of collaboration than in the partnership mode of collaboration. Japan is thus the exceptional case when it comes to the mode of co-authorship collaboration with the Philippines.

### 4.3.4 International co-authorships: Singapore
Scientific publications without international co-authors were dominant in Singapore until around the year 2000. From this point on, both BIRCs and MIRCs have grown steadily as shares of total scientific output. Papers without international collaborators made up just over one-third of total output in 2015 (Figure 10).

Figure 10 International co-authorships by type, Singapore 1980-2015

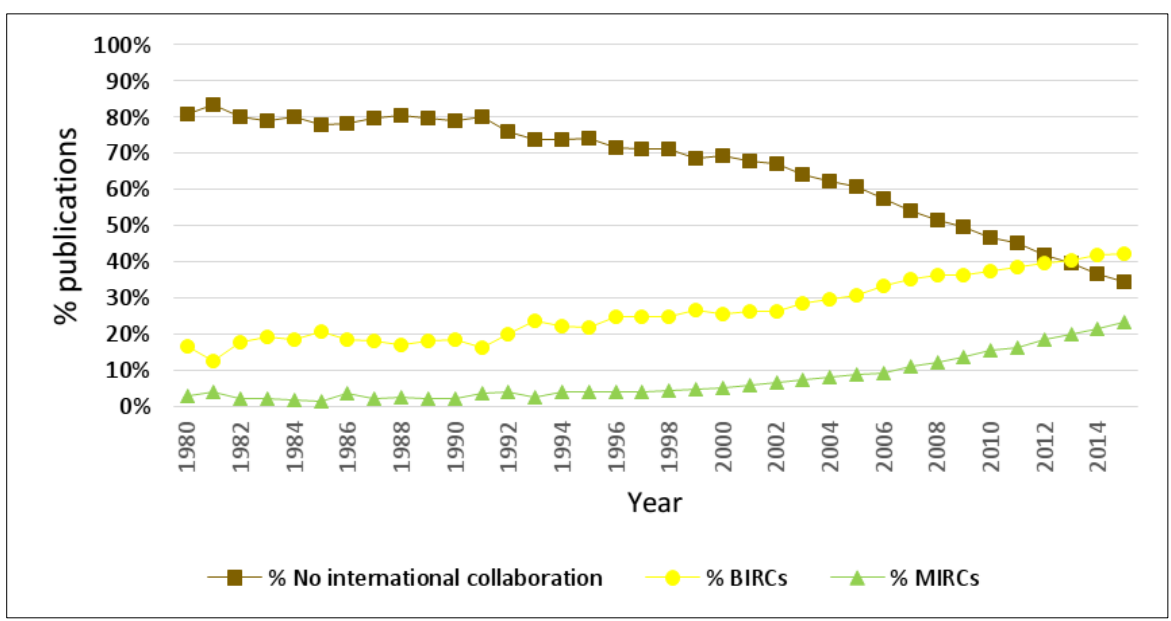

In 2015, BIRCs and MIRCs combined made up around 65% of Singapore's scientific output. In terms of co-authorship partners, the USA, China and the UK were the three most common locations of co-authors in the Singapore case (Table 5).

Table 5 International co-authorship countries, by type, Singapore 1980-2015

| SINGAPORE | | | |
|---|---|---|---|
| BIRCs | | MIRCs | |
| country | pubs | country | pubs |
| PR CHINA | 15363 | USA | 9820 |
| USA | 12149 | PR CHINA | 6433 |
| AUSTRALIA | 4012 | UK | 4696 |
| UK | 3764 | AUSTRALIA | 4182 |
| JAPAN | 1810 | GERMANY | 2552 |
| INDIA | 1541 | JAPAN | 2194 |
| CANADA | 1518 | CANADA | 2163 |
| GERMANY | 1095 | FRANCE | 1855 |
| SOUTH KOREA | 1078 | SOUTH KOREA | 1488 |
| TAIWAN | 981 | TAIWAN | 1367 |

Singapore-based authors' international co-authorships are strongly biased toward the partnership mode of collaboration, particularly with authors based in China and USA. Whilst researchers based in the USA and China are also the leading co-authors for network mode collaborations, the numbers are comparatively lower, particularly in the case of China collaborations. In contrast, other prominent collaborator locations, including the UK, Australia, Germany, Japan, Canada, France, South Korea and Taiwan, are all more commonly linked through the network mode of collaboration.

### 4.3.5 International co-authorships: Thailand

Papers with no international co-authors and BIRCs made up similar shares of the total output of Thailand from the early 1990s until around 2006. Since that time papers with no international co-authors have increased marginally to around half of all output. At the same time MIRCs have increased as a share of output, while BIRCs have declined slightly. MIRCs have been increasing as a share of total output slowly, but steadily, across the entire period 1980-2015 (Figure 11).

Figure 11 International co-authorships by type, Thailand 1980-2015

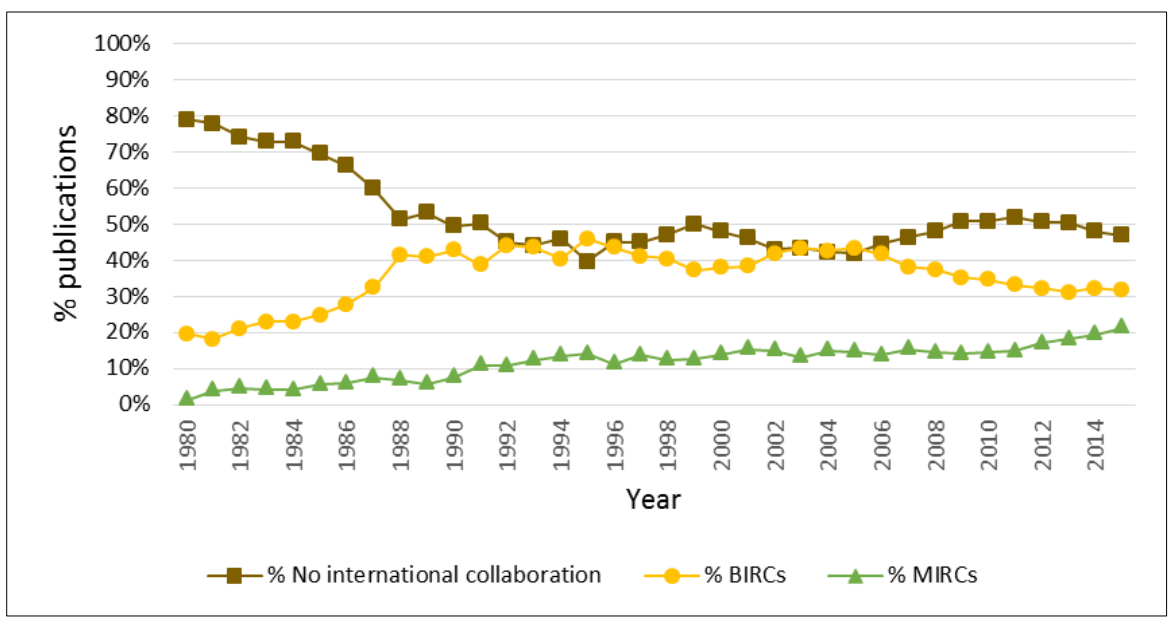

BIRCs comprised around 31% of total output from Thailand in 2015, with MIRCs making up a further 20%. In terms of co-authorship partners, the USA, Japan and the UK were the three most common locations of co-authors in the Thailand case (Table 6).

Table 6 International co-authorship countries, by type, Thailand 1980-2015

| THAILAND | | | |
|---|---|---|---|
| BIRCs | | MIRCs | |
| Country | pubs | country | pubs |
| USA | 8312 | USA | 5252 |
| JAPAN | 5996 | UK | 3394 |
| UK | 2417 | PR CHINA | 2405 |
| AUSTRALIA | 1884 | AUSTRALIA | 2243 |
| GERMANY | 971 | JAPAN | 2215 |
| CANADA | 833 | FRANCE | 1778 |
| FRANCE | 830 | GERMANY | 1536 |
| PR CHINA | 721 | INDIA | 1333 |
| MALAYSIA | 508 | SOUTH KOREA | 1247 |
| AUSTRIA | 432 | MALAYSIA | 1222 |

The USA is the main location of co-authors for both partnership mode and network mode collaborations of Thailand-based authors, with Japan second for the partnership mode and prominent also for the network mode. Both USA and Japan co-authorships are strongly biased toward the partnership mode of collaboration. Other prominent collaborator countries, including the UK, Australia, China, France, Germany, South Korea and Malaysia, are all biased toward the network mode of collaboration.

### 4.3.6 International co-authorships: Vietnam

In the early 1990s, BIRCs became the predominant way of publishing scientific papers for Vietnam-based authors. MIRCs and papers with no international collaboration have maintained a similar share of total output for the last twenty years, although MIRCs appear to be continuing to grow slowly as a share of output in recent years (Figure 12).

Figure 12 International co-authorships by type, Vietnam 1980-2015

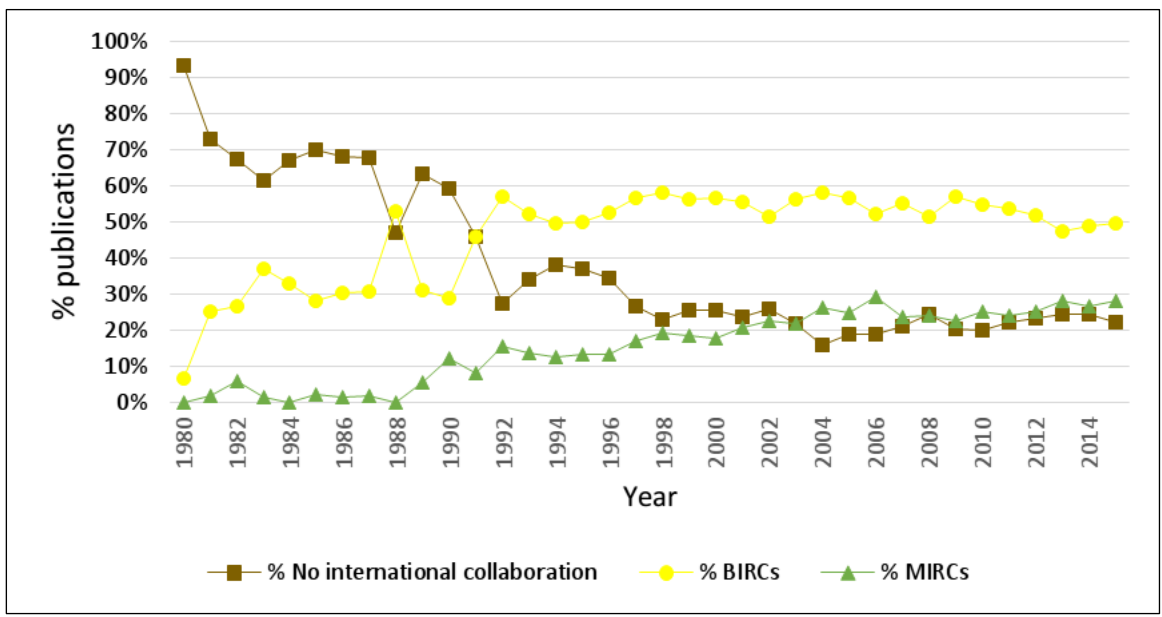

BIRCs comprised around 50% of total output from Vietnam in 2015, with MIRCs making up slightly less than 30%. In terms of co-authorship partners, the USA, Japan and France were the three most common locations of co-authors in the Vietnam case (Table 7).

Table 7 International co-authorship countries, by type, Vietnam 1980-2015

| VIETNAM | | | |
|---|---|---|---|
| BIRCs | | MIRCs | |
| country | pubs | country | pubs |
| JAPAN | 1567 | USA | 1715 |
| SOUTH KOREA | 1320 | UK | 1325 |
| FRANCE | 1137 | FRANCE | 1088 |
| USA | 917 | PR CHINA | 922 |
| AUSTRALIA | 764 | GERMANY | 897 |
| GERMANY | 684 | JAPAN | 889 |
| BELGIUM | 501 | THAILAND | 710 |

| NETHERLANDS | 430 | AUSTRALIA | 681 |
|---|---|---|---|
| SWEDEN | 391 | SOUTH KOREA | 674 |
| UK | 380 | NETHERLANDS | 651 |

The two largest collaborator locations for international co-authorships with Vietnam-based authors are the USA and Japan. Japan and South Korea are the major locations for partnership mode collaborations, while the USA and UK are the major location for network mode collaborations. Japan and South Korea are strongly biased toward the partnership mode of collaboration with Vietnam, while the USA and the UK are strongly biased toward the network mode. Of other prominent collaborators, co-authorships with France-based authors are relatively balanced between the two modes of collaboration, while China-based authors are concentrated in the network mode.

### 4.3.7 International co-authorships of the ASEAN6: Summary

This section has reviewed the international co-authorship patterns of the six largest knowledge producing countries among the ASEAN group of nations. It has established that there is considerable variation in the co-authorship patterns of the ASEAN6. There is some diversity in the degree to which each country partners with international co-authors as a share of the total scientific output. There is considerable diversity with the international co-authorships in terms of whether they are collaborations with authors from just one other country (BIRCs) or collaborations with authors from two or more countries (MIRCs).

There is also some diversity in the most prominent co-author locations for each country. However, an overall dominance in this regard exists in relation to the USA and Japan. A second rank of prominent collaborator locations for co-authorships with the ASEAN6 includes Australia, China and the UK. There is also very considerable diversity in the mode of collaboration that characterizes co-authorships with these main collaborator countries. The USA has co-authorships biased toward the network mode of collaboration with Indonesia, Malaysia and Vietnam, and the partnership mode with Singapore and Thailand. In contrast, Japan has co-authorships biased toward the partnership mode with all of the ASEAN6 except Singapore, which is slightly balanced toward the network mode. Australia-based co-authors are relatively even divided between partnership and network modes of collaboration, while UK-based authors are more strongly involved in the network mode. China is very different, being less prominent and mainly involved in network mode collaborations with several ASEAN6 countries, but being one of Singapore's two main collaborator locations and with a strong emphasis on the partnership mode of collaboration.

Overall, whilst there are some strong similarities in the modes and locations of international co-authorships between several of the ASEAN6, there are also unique features in each country case. Some potential broad explanations for the variation uncovered are outlined in the Discussion section of the paper. However, prior to that, the following section summarises data on the role of scientific fields and specialisations in shaping the scientific co-authorships of the ASEAN6.

## 4.4 Scientific fields of ASEAN6 international co-authorships

This section summarises the distribution of the international co-authorships of the ASEAN6 by scientific subject categories. For each ASEAN6 country we look at the ten subject categories with most co-authored publications and divide these co-authorships by the mode of collaboration. This section is intended to build information about the relevance of scientific specialisation for understanding the international co-authorships of the ASEAN6 and, as such, produces only very initial indications.

The results are illustrated in bar charts which show the BIRCs/MIRCs proportion and the raw number of MIRCs publications (to give an indication of subject area scale) for each subject area. The subject areas are ordered with the horizontal bar representing the subject area with the most international co-authored publications at the bottom of each country chart (Figure 13).

Figure 13 International co-authorships by collaboration type, top ten subject categories, ASEAN6 countries, 1980-2015

Indonesia

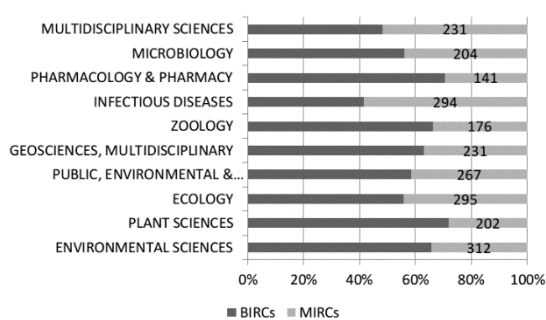

Malaysia

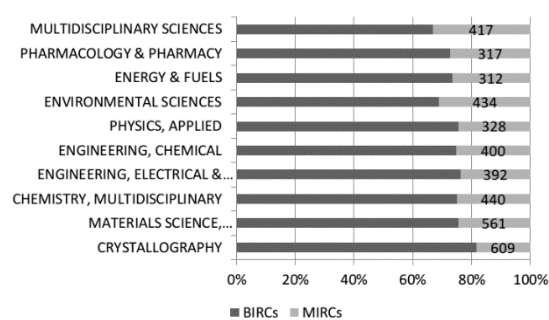

Philippines

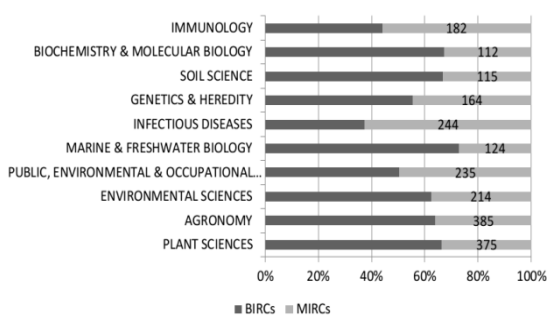

Singapore

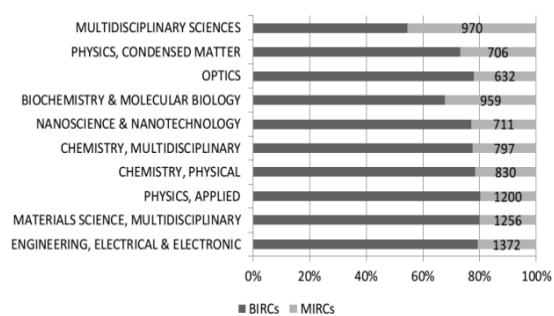

Thailand

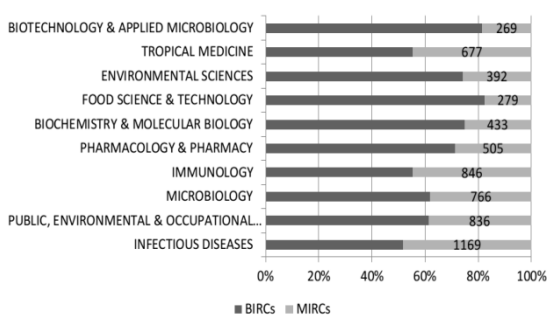

Vietnam

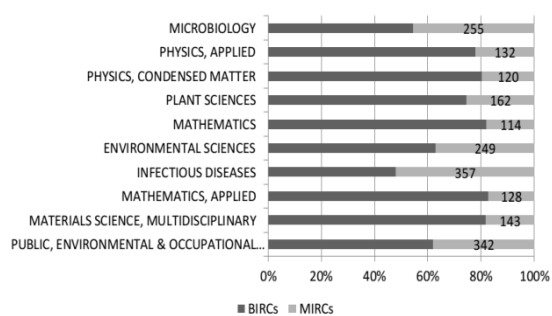

In terms of variation between ASEAN6 countries by major subject areas for international co-authorships it is evident that each country has a somewhat unique profile. The subject area with the most international co-authorships (bottom row of each country chart) is different in all countries. Environmental science is a top ten subject area in all countries except Singapore, while optics only features in the Singapore top ten. In general, Indonesia, the Philippines and Thailand are focused on natural sciences, plant sciences and the environment, along with medical and disease subject areas. Malaysia and Singapore have a stronger focus on engineering, materials science, physics and electronics areas. Vietnam mixes some of the elements with a focus on mathematics and physics subject areas.

In terms of variation between countries by collaboration type, Malaysia and Singapore are relatively more strongly involved in the partnership (BIRCs) mode of collaboration in the subject areas where they have high international co-authorship levels. Indonesia and the Philippines have the most diverse mix between partnership and network (MIRCs) modes of collaboration, while Thailand and Vietnam fall between these other two pairings of countries.

Looking at the distribution of co-authorships by collaboration type for individual subject areas, we can see that environmental science, which is common to all top ten lists apart from Singapore, has a relatively similar distribution of co-authorships by collaboration type across the five other countries. Comparing infectious diseases and plant science subject areas, across four and three countries respectively, a relatively consistent distribution of co-authorships by collaboration type is also evident.

In summary, this brief look at the international co-authorships of the ASEAN6 by scientific subject areas has shown diversity among the countries in terms of the modes of collaboration supporting their areas of scientific strength. Interestingly, an ad hoc comparison of three subject areas that are prominent in multiple countries showed quite consistent distributions of co-authorships by type across these different countries. This would suggest that the influence of scientific specialisation on the structure of international co-authorships requires further investigation.

# 5. Discussion

This article started out from the question of whether bi-lateral international co-authorships (BIRCs) and multi-lateral international co-authorships (MIRCs) contribute differently to the structuring of global research networks. The evidence shown, at both the level of the global system and the ASEAN sub-system, supports a positive response to this question. The evidence is that MIRCs are not only constructed by adding together countries already involved in BIRCs. Rather, the shift to the MIRCs mode of co-authorship appears to lead to the inclusion of additional countries in significant international collaborations and levels of scientific production. The remainder of the paper discusses the dynamics observed that may be contributing to this picture and offers some initial explanations for some of these phenomena.

There is no doubt that international scientific collaboration, when measured by the proxy of international co-authorship of scientific articles, is increasing. In 2015, international co-authorships constituted 60 per cent of scientific output from the ASEAN6. As a share of those co-authorships, MIRCs, or what we call the networked mode of collaboration, has been growing more rapidly than BIRCs, which we call the partnership mode of collaboration, in recent years at the global level. In the case of the ASEAN6 sub-system, the rate of growth in MIRCs has been considerably faster. The ratio of MIRCs to total scientific output for the ASEAN6 sub-system in 2015 was approximately 20%. This is more than three times the ratio of the global system. At the same time the ratio of BIRCs to total output in the ASEAN6 sub-system (40%) was double the ratio for the global system.

Clearly international collaboration has been, and continues to be, an important element of the development of the science and research systems in the ASEAN6 countries. From the evidence presented in this paper we cannot offer any definitive interpretation about the relative importance of BIRCs and MIRCs to this process. The historical pattern is largely consistent across the ASEAN6 however, with BIRCs accounting for more international co-authorships than MIRCs in the early part of the period observed (1980-2015), but with MIRCs constituting an ever increasing share in recent years. The transition to a greater proportion of MIRCs collaborations may have the potential to raise the quality of scientific outputs in terms of their citation impact, at least in some scientific fields (Wagner et al. 2017).

We considered two main factors in seeking to increase our understandings of the rates of participation in BIRCs and MIRCs we observed for each of the ASEAN6 countries, the co-authorship partner countries and scientific specialisation. In terms of co-author countries, the two most consistent partner countries for ASEAN6 authors are the USA and Japan. There are notable exceptions to this including the involvement of Malaysia in co-authorships with the UK and Singapore with China. However, comparing the USA and Japan there is an apparent difference in the prevailing collaboration mode. In the case of the USA, in raw numbers MIRCs is the more common collaboration mode in relation to all countries except Singapore and Thailand. In the case of Japan, BIRCs are the more common mode in relation to all countries except Singapore.

The USA and Japan thus appear to have differing collaboration styles when it comes to their linkages to the ASEAN6 countries. The USA is implicated in a more multiple networked model of collaboration in most countries, while Japan is involved in a narrower two-partner model. There are a number of factors that may help to explain this apparent difference. First, the USA has traditionally been the preferred destination for PhD and post-doctor researchers from the Asia-Pacific region (Woolley et al. 2008). The opportunity to meet and interact with colleagues not just from the US, but also from other international countries, is a feature of the US system of scientific training. The networked model of collaboration involving the USA may partly reflect professional ties built within the diverse internationalized environment of US universities.

Second, a significant proportion of foreign PhDs and post-docs who train in the USA remain in the USA afterwards to work (Lee 2004). Some become permanent migrants. As the work of Meyer and colleagues has shown, such 'migrants make networks' opening pathways for the next generation of students to visit and train overseas. Such diaspora knowledge networks provide spaces of circulation that can reinforce the exposure of students from the ASEAN6 to internationalized professional networks. Particularly until more recent years, the USA was a much more likely destination for mobile and migrant scientists than Japan. Third, the commonality of English as the dominant language in the USA and for science publishing may be another reinforcing factor.

There are undoubtedly other factors that may influence the apparent difference in collaboration styles between the USA and Japan with the ASEAN6. For example, institutional arrangements or rules related to research and/or project funding may be relevant. Laws and rules related to rights to residence, access to labour markets, and conversion to citizenship, may have an effect on the relative diversity of the research community in the USA and Japan, therefore providing more or less diaspora contact points for authors based in the ASEAN6 countries. Overall, these potential explanations provide motivation for further investigation into the factors shaping the apparent differences in the styles of these major collaborators with the ASEAN6.

It was noticeable that Singapore is an exception to the predominant mode of collaboration for both the US and Japan. In the case of the US, the strong partnership mode of collaboration is no doubt part of a very direct scientific relationship, but it may also be linked to the nature of the science being conducted in Singapore when compared to the rest of the ASEAN6. In the case of Japan, the dominance of China in so-authorships with Singapore, alongside the US, appears to push Japan into the unfamiliar position of being more predominantly a network member than an exclusive partner. Singapore, along with Malaysia and Thailand, has produced more papers through MIRCs than BIRCs in the period 1980-2015 (Table A). In this paper there is no space to go more deeply into the collaboration styles of the ASEAN6 and their major collaborators, but explanations for these different mixes of collaboration styles remains a topic deserving further research.

Field differences in the structure of international co-authorships were also investigated in a limited way. It was apparent that ASEAN6 countries have some similar and some different subject areas of publishing strength, which are produced through different emphases on BIRCs and MIRCs. However, a simple comparison of the balance of BIRCs and MIRCs in producing papers in specific subject areas showed that this balance was relatively consistent regardless of the individual country. On this very preliminary indication, subject areas may have relatively consistent partnership (BIRCs) to network (MIRCs) collaboration ratios. Possible explanations for this may be linked to degrees of interdisciplinarity in particular areas. It seems possible that the networked mode of collaboration could be an effect of the need to connect different disciplinary knowledges together. A different explanation for a greater emphasis on the networked mode of collaboration could also be linked to key infrastructures, in which research can only be conducted via access to key equipment available only in certain countries – potentially bringing in additional co-authors on scientific outputs using these facilities.

In terms of the broad question of how the structure of international co-authorships may impact on the development of science and research systems, this research provides a mixed insights. The evidence of the comparison between the structure of the BIRCs and MIRCs co-authorship networks of the ASEAN6 suggests that the network mode of collaboration leads to higher levels of scientific productivity for an extended number of countries. The role of the ASEAN6 countries as intermediaries to additional, often smaller, countries can be seen in the visualizations of the MIRCs network. These networked relationships likely leverage the partnerships ASEAN6 countries have with scientific research powerhouses such as the USA and Japan, helping to diffuse the benefits of these relationships more widely. However, at the same time this evidence also raises a question regarding the rhetoric supporting the universal benefits of global networks in science. The map of the BIRCs network shows a more exclusive version of the global science collaboration process. The extent to which the functioning of the global scientific network is actually determined by two country partnerships requires further and much deeper investigation. Ultimately, understanding the different effects of the partnership and network modes of collaboration within the global landscape of science could have significant implications for policy and institutional arrangements, particularly where development and capacity building objectives are priorities.

Future research with these data will focus on two main avenues. First, the impact of scientific fields on the structuring of cooperation into partnership and network modes will be explored. It is likely that the epistemic properties of scientific fields are more important for the structuring of global networks than national scientific research systems in some sciences. Second, comparative analyses offer another way to explore the dynamics which may be driving the results found in this paper. An analysis of the patterns and structure of other developing regions' co-authorship networks, incorporating a more developed analysis of the importance of scientific fields, will seek to extend this exploratory study toward some conclusive findings.